\documentclass[aip,graphicx]{revtex4-1}

\usepackage[utf8]{inputenc}
\usepackage[pdftex]{graphicx}
\usepackage{amsmath,amsfonts}
\usepackage[amssymb,Gray]{SIunits}
\usepackage{cancel}

\newcommand{\rmd}{\ensuremath{\mathrm{d}}}
\newcommand{\abs}[1]{\ensuremath{\left\vert #1 \right\vert}}
\newcommand{\erw}[1]{\ensuremath{\langle #1 \rangle}}

\newcommand{\bra}[1]{\ensuremath{\langle #1 \hspace{-2pt} \mid}}
\newcommand{\ket}[1]{\ensuremath{\mid \hspace{-1pt} #1 \rangle}}
\newcommand{\braket}[2]{\ensuremath{\langle #1 \hspace{-2pt} \mid \hspace{-1pt} #2 \rangle}}
\newcommand{\R}{\mathbb{R}}

\newcommand{\Tr}{\mathrm{Tr}}

\newcommand{\schr}{Schr{\"o}\-din\-ger}

\begin{document}


\title{Three little paradoxes: making sense of semiclassical gravity} 



\author{André Großardt}
\email[]{andre.grossardt@uni-jena.de}
\affiliation{Institute for Theoretical Physics, University of Jena, Germany}


\date{\today}

\begin{abstract}
I review the arguments most often raised against a fundamental coupling of classical spacetime to quantum matter. I show that an experiment by Page and Geilker does not exclude such a semiclassical theory but mandates an inclusion of an objective mechanism for wave function collapse. In this regard, I present a classification of semiclassical models defined by the way in which the wave function collapse is introduced. Two related types of paradoxes that have been discussed in the context of the necessity to quantize the gravitational field can be shown to not constrain the possibility of a semiclassical coupling. A third paradox, the possibility to signal faster than light via semiclassical gravity, is demonstrably avoided if certain conditions are met by the associated wave function collapse mechanism. In conclusion, all currently discussed models of semiclassical gravity can be made consistent with observation. Their internal theoretical consistency remains an open question.
\end{abstract}

\pacs{}

\maketitle 

\section{Introduction}

The quantization of the gravitational interaction has crystallized as a considerably more challenging endeavor than the quantization of matter fields. Paired with the success of classical general relativity when it comes to the accuracy of predictions, the question has been raised whether we are on the right track, or whether the question \emph{how} to quantize gravity is misguided and a semiclassical theory, in which only matter is quantized and spacetime remains fundamentally classical, should be sought for instead.

The discussion about whether or not gravity must be quantized reaches back as far as to the early days of quantum field theory. At the 1957 Chapel Hill Conference\cite{dewittRoleGravitationPhysics1957} Feynman famously introduced a thought experiment in which a spin superposition state becomes entangled with the position of a macroscopic mass, allegedly showing the need for a quantized gravitational field---at least if one is willing to ``believe in quantum mechanics up to any level''\cite{dewittRoleGravitationPhysics1957}. Similar thought experiments\cite{eppleyNecessityQuantizingGravitational1977,kibbleSemiClassicalTheoryGravity1981,pageIndirectEvidenceQuantum1981,baymTwoslitDiffractionHighly2009,belenchiaQuantumSuperpositionMassive2018} have been repeatedly brought into the discussion since, to prove the necessity of quantization, with a direct refutation\cite{kibbleSemiClassicalTheoryGravity1981,huggettWhyQuantizeGravity2001,mattinglyQuantumGravityNecessary2005,mattinglyWhyEppleyHannah2006,kieferQuantumGravity2007,albersMeasurementAnalysisQuantum2008,kentSimpleRefutationEppley2018,rydvingGedankenExperimentsCompel2021} often following on the heels.
As recently as last summer, the question has been asked: \textit{``Do Gedankenexperiments compel quantization of gravity?''}\cite{rydvingGedankenExperimentsCompel2021} with the answer still the same as much more humbly and eloquently\footnote{With the greatest appreciation, I welcome any clues regarding the mode of operation of the mysterious ``Chicago machine'' transforming hogs into sausages.} stated by Rosenfeld\cite{rosenfeldQuantizationFields1963} 59 years ago: \textbf{No!}

I take a closer look at these thought experiments and consistency arguments and assert that the underlying concepts of semiclassical gravity fall into three categories, ultimately related to their incorporation of the measurement process. I first analyze the implications of the sole experiment on semiclassical gravity that has actually been conducted\cite{pageIndirectEvidenceQuantum1981}, concluding that it only rules out a rather specific sub-category of semiclassical theories: the Everettian ones which are based on the quantum state without ever collapsing the wave function. I further discuss the three classes of paradoxes that are brought up as arguments against semiclassical gravity, and explain why only one of the paradoxes considered in connection with semiclassical gravity---and only in the third case of the traditional, $\psi$-ontic semiclassical models---poses a challenge for the coupling of classical gravity to quantum matter. In order to resolve it, a semiclassical model for gravity must be of a twofold nature: providing both the coupling of quantum matter to classical spacetime and a dynamical description of wave function collapse.

\section{Models of semiclassical gravity}\label{sec:models}

Before we begin, let us clarify the definition of some notions. Talking about the necessity of quantizing gravity, one faces the obvious question what it means to \emph{quantize} gravity. For the purpose of this work, quantized gravity refers to any model in which spacetime is \emph{not classical}. By contrast, we refer to a theory with classical spacetime satisfying Einstein's equations for \emph{some} right-hand side, as \textbf{semiclassical gravity}. In semiclassical gravity, matter is usually though of as being described by quantum fields on said classical curved spacetime, with some freedom of choice regarding the backreaction through Einstein's equations and---as we will see---the characterization of measurement.
The semiclassical Einstein equations\cite{mollerTheoriesRelativistesGravitation1962,rosenfeldQuantizationFields1963},
\begin{equation}\label{eqn:sce}
R_{\mu\nu} - \frac{1}{2} g_{\mu\nu} R = \frac{8 \pi G}{c^4} \bra{\Psi} \hat{T}_{\mu\nu} \ket{\Psi} \,,
\end{equation}
where the left-hand side is the Einstein tensor, constructed from the scalar and tensor curvatures $R$ and $R_{\mu\nu}$ as well as the metric $g_{\mu\nu}$, and the right-hand side contains the expectation value of the stress energy operator in the quantum state $\Psi$, are only one of potentially many possible realizations of semiclassical gravity. 

In order to be as precise as possible, I give a vague sketch of a definition by noting that a model of semiclassical gravity generally consists of (some of) the following ingredients:
\begin{enumerate}
 \item[(i)] a classical spacetime, i.\,e.\ a pseudo-Riemannian 4-manifold $(\mathcal{M},g)$ with metric $g_{\mu\nu}$,
 \item[(ii)] a set of quantum fields described by the total state $\ket{\Psi} \in \mathcal{H}$, where the state space $\mathcal{H}$ is constructed in the spirit of quantum fields on curved spacetime\cite{waldQuantumFieldTheory1994},
 \item[(iii)] a set of ``hidden variables'', i.\,e.\ classical fields $\Lambda : \mathcal{M} \to \R^n$ (which can be empty),
 \item[(iv)] the U-process\cite{penroseGravityRoleQuantum1996}, i.\,e.\ a dynamical law for the quantum states $\ket{\Psi}$, usually given in the form of a Lagrangian for the fields (e.\,g.\ the standard model of elementary particles),
 \item[(v)] the R-process\cite{penroseGravityRoleQuantum1996}, i.\,e.\ a dynamical law governing quantum state reduction during measurement-like situations,
 \item[(vi)] a classical stress-energy tensor $T_{\mu\nu}(\Psi,\Lambda,x)$ defining the right-hand side in the semiclassical Einstein equations, depending on the quantum state $\Psi$ and the hidden variables.
\end{enumerate}
Already the proper definition of the Hilbert space (ii) is nontrivial, as is understanding the dynamics (iv), not to mention the consistent inclusion of backreaction (vi). Nonetheless, the biggest question mark is attached to the definition of the R-process (v), of which we only know its effect, namely that measurements result---at least to good approximation---in eigenstates with Born rule probabilities. In the case that there are hidden variables, the R-process must include the dynamical laws for $\Lambda$, e.\,g.\ the guiding equation for particle coordinates in the de Broglie-Bohm theory\cite{bohmSuggestedInterpretationQuantum1952a}. Of course, U- and R-processes will generally not be strictly separable but only limiting cases of a joint dynamics for all degrees of freedom in the model: $\ket{\Psi}$, $\Lambda$, and $g_{\mu\nu}$. For the purpose of this article, I distinguish three classes of semiclassical models.

\paragraph{$\psi$-ontic semiclassical gravity} refers to all models in which $T_{\mu\nu}(\Psi,x)$ is a well defined function of the state $\ket{\Psi}$, independent of the hidden variables (regardless whether there are hidden variables governing the R-process or whether the set of hidden variables is empty). Nonrelativistically, some functional of the wave function plays the role of a mass density; e.\,g.\ $\rho(t,\vec r) = m \abs{\psi(t,\vec r)}^2$ for a single particle in the M\o{}ller-Rosenfeld model based on the semiclassical Einstein equations~\eqref{eqn:sce}.

\paragraph{Hidden variable semiclassical gravity} refers to models in which $T_{\mu\nu}(\Lambda,x)$ is primarily a function of the hidden variables (although it implicitly depends on the quantum state via the R-process). One might think that this occurrence of hidden variables $\Lambda$ is a mere philosophical peculiarity, and that any such model should reduce to a $\psi$-ontic one by including $\Lambda$ as an explicit function in the definition of the right-hand side $T_{\mu\nu}(\Psi,x)$. Nonetheless, the distinction is useful if it comes to the question of wave function collapse, as the existence of $\Lambda$ allows for a $\psi$-epistemic interpretation and only the dynamics of $\Lambda$, not those of the quantum state, must be compatible with principles of general relativity.

\paragraph{Stochastic semiclassical gravity,} finally, refers to models where the right-hand side of Einstein's equations depends on the density matrix $\hat{\rho}$, or rather on quantities derivable from $\hat{\rho}$ as the outcomes of local measurements. The U- and R-processes then determine $\hat{\rho}$ which allows for the usual stochastic interpretation in terms of Born rule probabilities for measurement outcomes.
With this definition, the stochastic models are clearly set apart from the $\psi$-ontic ones, in which the gravitational field can depend on properties of the state that in standard quantum mechanics are not measurable by any local measurement---specifically in nonlocally entangled states. Again, the difference is not obvious as long as one disregards wave function collapse and takes the traditional point of view in which density matrices are merely a statistical tool to keep track of the dynamics of an ensemble of pure states.
When it comes to the operational perspective, in which the density matrix plays the central role of describing physical reality whereas the wave function becomes a mere tool of bookkeeping of an observer's knowledge, there is, however, a crucial difference. This is also the case in collapse models\cite{bassiModelsWavefunctionCollapse2013}, where the density matrix still obeys a linear and deterministic dynamical law defined by a Lindblad type master equation, whereas the evolution of the wave function follows a stochastic differential equation.
In this case of a linear master equation, the density matrix based approach ensures compatibility with quantum mechanical predictions which avoids most paradoxes; it does however raise difficult questions of interpretation.

Note that the notions of $\psi$-ontic, hidden variable, and stochastic models of semiclassical gravity refer solely to the way in which quantum matter is coupled to classical spacetime. The definition of $T_{\mu\nu}$ notwithstanding, the quantum mechanical interpretation as such can be different. For instance, there could be hidden variables within a $\psi$-ontic model, or an objective collapse as part of the R-process in a semiclassical model based on hidden variables.

Needless to say, there are other possible definitions of the notions of both semiclassical and quantized gravity which may not agree with the ones adopted here---at least not for all models. For instance, it is often presumed that quantized gravity yields perturbative quantum gravity as its low energy limit; however, for quantized gravity as defined here, this is not a requirement. Similarly, recent proposals\cite{boseSpinEntanglementWitness2017,marlettoGravitationallyInducedEntanglement2017} attracted some attention, which suggest to detect entanglement generation via gravity. As far as the definitions used here are concerned, there is no conclusive argument that in the experimental scenarios at hand \emph{all} models of quantized gravity would result in a confirmative observation of entanglement, nor is there a compelling proof for separability of the respective equations of motion---i.\,e.\ no entanglement---in \emph{all} semiclassical models\cite{reginattoEntanglingQuantumFields2019,palExperimentalLocalisationQuantum2021,carneyNewtonEntanglementGraviton2021,donerGravitationalEntanglementEvidence}.

\subsection{Examples for semiclassical gravity models}
The go-to example of a $\psi$-ontic model is of course to source gravity via the expectation value of the stress energy operator according to the semiclassical Einstein equations. This model, independently proposed by M\o{}ller\cite{mollerTheoriesRelativistesGravitation1962} and Rosenfeld\cite{rosenfeldQuantizationFields1963}, has been studied extensively, especially in the nonrelativistic limit where it yields the Schrödinger-Newton equation\cite{diosiGravitationQuantummechanicalLocalization1984,penroseQuantumComputationEntanglement1998} and makes distinctive predictions\cite{giuliniGravitationallyInducedInhibitions2011,yangMacroscopicQuantumMechanics2013,grossardtOptomechanicalTestSchrodingerNewton2016,grossardtDephasingInhibitionSpin2021}.
In fact, requiring consistency with principles of general relativity and the correct classical limit puts tight constraints on the possible choices of stress-energy tensors\cite{waldQuantumFieldTheory1994}. Whether any other consistent models of this type can be defined is unclear.

Hidden variable models, despite being the main subject of inconsistency arguments, as I will argue below, are not commonly discussed. Nevertheless, at least in the Newtonian limit one can easily define such a model based on the de Broglie-Bohm theory\cite{bohmSuggestedInterpretationQuantum1952a}. There, one describes the motion of particles with coordinates $q$ in dependence of the wave function with a guiding equation $\dot{q} = f[\psi](q,t)$. One can then simply use the particle coordinate $q$ in order to source a Newtonian gravitational potential which enters the Schrödinger equation for the wave function $\psi$. Unfortunately, the naive relativistic generalization is inconsistent\cite{struyveSemiclassicalApproximationsBased2020}.

As far as the stochastic models are concerned, a fully general relativistic version of such a model has recently been presented by Oppenheim\cite{oppenheimPostquantumTheoryClassical2021}, not unlike the one introduced by Albers et al.\cite{albersMeasurementAnalysisQuantum2008} for scalar gravity. Rather than a single spacetime manifold, these models describe statistical ensembles of spacetimes. Therefore, strictly speaking, they do not qualify as models of semiclassical gravity as defined here, but could potentially be regarded as a theory of (semi-)classical statistical mechanics for such models.

On the other hand, one can obtain a genuine semiclassical model, which can also be endowed with a $\psi$-ontic interpretation, starting from collapse models.\cite{tilloySourcingSemiclassicalGravity2016} The stochastic collapse of the wave function renders the evolution of the density matrix linear, allowing to calculate the ``signal'' of the mass distribution in analogy to weak measurements. This signal is then fed back into Einstein's equations as a source of spacetime curvature. At least that is the idea; so far only a nonrelativistic version has been constructed as the consistent definition of relativistic collapse models\cite{bedinghamCollapseModelsRelativity2020} poses serious problems.

The often cited model by Kafri et al.\cite{kafriClassicalChannelModel2014} to describe the Newtonian interaction between two masses by local operations and classical communication can be considered a prototype of the Tilloy-Di\'osi collapse based model\cite{tilloySourcingSemiclassicalGravity2016}. By itself, it does not qualify as a model for semiclassical gravity as per the definition here, as it does not provide a meaningful notion of classical spacetime.

\section{Refutation of Everettian semiclassical gravity}\label{sec:rho}

When people argue for the necessity to quantize the gravitational field, one of the most common sources in which they find support is a letter by Page and Geilker\cite{pageIndirectEvidenceQuantum1981}. One of the more interesting facts about this work is that the results of what is best described as a student lab experiment got published by no lesser than the Physical Review Letters. This is even more astonishing, considering that their experiment rules out only a very specific case of stochastic semiclassical models: the Everettian one in which the wave function never collapses and the semiclassical Einstein equations~\eqref{eqn:sce} are used to source spacetime curvature. As there is no R-process but an entirely unitary dynamics, the equivalence between the pure state density matrix $\hat{\rho} = \ket{\Psi}\bra{\Psi}$ and the global state $\ket{\Psi}$ of all matter fields allows to interpret this model also as a special case of the $\psi$-ontic models.

The advances that came with the formalism of quantum mechanics as a statistical theory, based on the density matrix of a system and its evolution via a linear master equation, have resulted in a popular point of view from which quantum mechanics is fundamentally a stochastic theory, with the density matrix playing the central role and the wave function merely being a sometimes helpful tool. Nonetheless, the way the density matrix is introduced in quantum mechanics courses is mostly still based---at least implicitly---on the $\psi$-ontic view of pure Hilbert space states taking the fundamental role of describing physical reality and the density matrix representing stochastic ensembles of such pure states.

With regard to semiclassical gravity, and specifically the semiclassical Einstein equations~\eqref{eqn:sce}, this raises the question how to deal with different mixtures that represent the same probability distribution. For example, the density matrix for a superposition
\begin{equation}\label{eqn:superpos}
\ket{\psi_0} = \frac{1}{\sqrt{2}} \left( \ket{x_1} + \ket{x_2} \right)
\end{equation}
of a massive particle in two positions $x_1$, $x_2$, represented in the Hilbert subspace basis $\{\ket{x_1},\ket{x_2}\}$, will decohere like
\begin{equation}\label{eqn:density-decoherence}
 \hat{\rho}_0 = \ket{\psi_0} \bra{\psi_0}
 = \frac{1}{2}\begin{pmatrix} 1&1\\1&1 \end{pmatrix}
 \quad\longrightarrow\quad
 \hat{\rho}_t \approx \frac{1}{2}\begin{pmatrix} 1&0\\0&1 \end{pmatrix} \,,
\end{equation}
when coupled with some environment for a sufficient time $t$. In the nonrelativistic limit, the stress energy operator reduces to the mass density operator $\hat{T}_{\mu\nu} \approx \hat{m}(x) = m \ket{x}\bra{x}$. Taking the expectation value from the stochastic states then results in the same mass distribution $\rho(x) = \Tr \hat{\rho}_0 \hat{m}(x) = \Tr \hat{\rho}_t \hat{m}(x)$ for the initial and decohered states. Regardless of decoherence, according to equation~\eqref{eqn:sce} the superposition would gravitate like an equal distribution of half the total mass at both positions. This would be in obvious contradiction to observations in many everyday situations. In case there were any doubts left, it has also been experimentally ruled out by Page and Geilker\cite{pageIndirectEvidenceQuantum1981}.

However, only this specific version of semiclassical gravity where the uncollapsed global state, i.\,e.\ the density matrix for subsystems excluding the environment, acts as the gravitational mass density, is refuted by experiment. In this no-collapse situation, there is also no physical difference between the three categories of semiclassical gravity models introduced in the previous section. Disregarding the R-process, gravitational source terms based on the wave function or the density matrix can be substituted with each other, and the dependence on hidden variables becomes trivial.

If, on the other hand, $\hat{\rho}_t$ is understood as a mixture of the pure classical states $\ket{x_1}$, $\ket{x_2}$ instead, one would not expect any deviation of the gravitational field from that of a classical point mass and the outcome of the experiment becomes trivial.
This is the reason why Page and Geilker's argument applies neither to the hidden variable nor the $\psi$-ontic models (in the nontrivial case with an R-process): equation \eqref{eqn:density-decoherence} merely describes the ensemble of possible states but semiclassical gravity is sourced by the concrete representative in said ensemble.
In this case, in order to end up with such a mixture starting with the superposition state \eqref{eqn:superpos}, the entire state, including the environment, must undergo a nonlinear evolution (``collapse'')
\begin{equation}\label{eqn:collapse}
 \ket{\Psi}_0 = \ket{\psi_0} \otimes \ket{\text{env.}}
 \quad\longrightarrow\quad
 \ket{\Psi}_{t,i} = \ket{x_i} \otimes \ket{\text{env. for particle at } x_i}
\end{equation}
with probabilities given by the Born rule $P_i = \abs{\braket{x_i}{\psi_0}}^2$.
The right-hand side in Einstein's equations must be compatible with the collapse dynamics described by equation~\eqref{eqn:collapse} in the nonrelativistic limit. At the same time, the continuity equation $\nabla_\mu {T^\mu}_\nu = 0$, which follows from the vanishing of the covariant divergence of the Einstein tensor, must be obeyed. This condition puts strong constraints on the dynamical laws underlying the wave function collapse which are usually not satisfied by nonrelativistic collapse models.\footnote{An insightful discussion of the issue of energy conservation in semiclassical gravity is presented by Maudlin et al.\cite{maudlinStatusConservationLaws2020}, as was kindly revealed to me in a prudent referee report.}

One could also justify an agnostic view of collapse, as it is accepted in most text book formulations of quantum mechanics: the state of a system does not obey the \schr\ equation during a ``measurement''. Instead, one simply \emph{postulates} the state after measurement, and the \schr\ evolution law takes on again thereafter. Analogously, postulating a collapse according to equation~\eqref{eqn:collapse} and requiring the semiclassical Einstein equations~\eqref{eqn:sce} to hold before and after but not during the collapse\cite{okonWeightCollapseDynamical2018,maudlinStatusConservationLaws2020} would circumvent the issue. Such models can certainly be considered as ``semiclassical'' in some meaning of the word, although they are not semiclassical according to the definition given before which explicitly required the validity of Einstein's equations for some right-hand side.

We conclude that a consistent deterministic theory of semiclassical gravity must achieve both: provide a description how the right-hand side of Einstein's equations is determined for quantum matter \emph{and} a description how the coupling to a macroscopic system results in a nonlinear dynamics which produces quasi-classical pure states with Born rule probabilities. Notably, the inclusion of the wave function collapse also clarifies the outcome in the thought experiment proposed by Feynman\cite{dewittRoleGravitationPhysics1957}. Kibble\cite{kibbleSemiClassicalTheoryGravity1981}, who introduced a similar though experiment, already points out this connection between semiclassical gravity and measurement theory.

Similar thoughts apply to the stochastic models, except for the reasoning being reversed. Although equation~\eqref{eqn:density-decoherence} does apply for these models, contradictions with the experiment can be excluded by modifying the right-hand side in Einstein's equations. In the case of the collapse based model by Tilloy and Diósi\cite{tilloySourcingSemiclassicalGravity2016}, for instance, spacetime curvature is sourced by the signal $\erw{\hat{m}(x)} + \delta m(x)$ with some noise $\delta m$. The stochastic collapse of the wave function results in a gravitational field compatible with the actual measurement outcome, despite the density matrix still having the shape of equation~\eqref{eqn:density-decoherence}.

\section{Paradoxes of hidden variable semiclassical gravity}\label{sec:hidden}

With the conclusion of the previous section, that semiclassical gravity needs to be accompanied by a description of measurement, we are left with two consistent possibilities, depending on the answer to the question whether or not gravity ``can be used [...] to `collapse the wave function [...]'{}''\cite{eppleyNecessityQuantizingGravitational1977}.
With the ability to collapse the wave function comes the capability to acquire which-path information. 

The concept of which-path information must implicitly assume this information to be about \emph{something} more than the wave function, which does not contain any information about which of the possible states a system will collapse into. Hence, it is evident that the $\psi$-ontic point of view does not allow to acquire which-path information through gravitational observations. Instead, the gravitational field will contain information about the wave function in $\psi$-ontic models.
In hidden variable models, on the other hand, the common degree of freedom $\Lambda$ determines both the gravitational interaction and the outcome of wave function collapse. These models, therefore, clearly allow for the acquisition of which-path information through the gravitational interaction.

There are two types of paradoxes based on the acquisition of which-path information that have been discussed. Due to the above considerations, these do not pose any threat to the $\psi$-ontic models, whereas their relevance for the stochastic ones seems to depend somewhat on the concrete realization. Be that as it may, I will show that even in the case of the hidden variable models these paradoxes are easily resolved and pose no constraints on the set of possible models for semiclassical gravity.

Note that for the subsequent discussion I use Planck units with $G = c = \hbar = 1$. 

\subsection{Violation of position-momentum uncertainty relation}\label{sec:suba}

Assume we could scatter a classical gravitational wave off a quantum particle. As classical waves are not required to obey the de Broglie relation between wave length and momentum, we can choose a wave with $\lambda \ll 1/p$. If the deflection angle of this wave can be detected with sufficient precision, one can conclude the position of the particle with negligible change of its momentum, thereby violating the uncertainty relation $\Delta x \, \Delta p > 1$. This has been presented by Eppley and Hannah\cite{eppleyNecessityQuantizingGravitational1977} as an argument for the necessity of quantizing the gravitational field---and has been refuted many times. Huggett and Callender\cite{huggettWhyQuantizeGravity2001} as well as Kiefer\cite{kieferQuantumGravity2007} discuss the implications of Eppley and Hannah's thought experiment and the necessity to quantize the gravitational field in great detail, whereas Albers et al.\cite{albersMeasurementAnalysisQuantum2008} give an explicit counter-example for a consistent hybrid quantum-classical theory (scalar gravity with a quantized scalar field) and argue that even in a hybrid theory uncertainty of the quantum observables induces uncertainty on the classical ones. Kent\cite{kentSimpleRefutationEppley2018}, on the other hand, presents a simple refutation of the second aspect of Eppley and Hannah's argument, namely that scattering of a gravitational wave off the wave function would result in the problems with causality to be addressed in section~\ref{sec:psi}. For the discussion here, I focus on the objections raised by Mattingly\cite{mattinglyQuantumGravityNecessary2005,mattinglyWhyEppleyHannah2006}, who points out that, at least with the parameters given by Eppley and Hannah, there are some experimental obstacles hard to overcome even in principle---not least that for the given values their detector would lie within a black hole---and in any case, ``it may be that the uncertainty relations \emph{can} be violated [because] they haven't really been tested in this way.''\cite{mattinglyQuantumGravityNecessary2005} In combining those two lines of thoughts, one can repeat Mattingly's analysis in a slightly more general way, not only applying to the specific parameters chosen by Eppley and Hannah.

Digging into the details of the detection procedure outlined by Eppley and Hannah, they first describe the generation of a gravitational wave pulse by the collision of two massive objects of size $\lambda$ with a kinetic energy $E$. The wave being scattered by the particle of mass $m$ at distance $r$ from the generation event carries the energy $E_\text{sc} \sim E^2 m^2 \lambda^{-1} r^{-2}$. By comparing the energy density from the scattered gravitational wave in a distance $R$ from the particle with the local gravitational energy density one finds that the amplitude of the gravitational wave at the detector can be expressed as $A \sim E m R^{-1} r^{-1}$. Between the two ends of an oscillator of size $2L \lesssim \lambda$, mass $M$, and frequency $\omega_0 \ll \omega$, this induces the differential force\cite{misnerGravitation1973}
$F(t) = m \omega^2 L A \sin\omega t$, where we denote by $\omega = 2 \pi / \lambda$ the frequency of the wave. The result is a driven oscillation with frequency $\omega$, amplitude $L A$, and an oscillation energy $E_\text{osc} \sim M \omega^2 L^2 A^2 \sim M m^2 L^2 E^2 \lambda^{-2} R^{-2} r^{-2}$. The transition probability for such an oscillator is of the order of $E_\text{osc} / \omega_0$, implying that for detection one needs $N \sim \omega_0 / E_\text{osc}$ detectors with a total mass
\begin{equation}
 M_\text{tot} \sim \frac{M \omega_0}{E_\text{osc}}
 \sim \frac{\omega_0 \lambda^2 R^2 r^2}{m^2 L^2 E^2}
 \gtrsim \frac{\omega_0 R^2 r^2}{m^2 E^2} \gtrsim \frac{\omega_0 R^2}{m^2} \,.
\end{equation}
where we require $E \lesssim r$ in the final step, as otherwise our particle would vanish in the singularity created during the generation of the gravitational wave.

Eppley and Hannah argue that despite the proposed low value of $\omega_0$, the time of measurement can be made short because it suffices to detect whether the energy of one of the oscillators increased by $\omega_0$. However, as Mattingly\cite{mattinglyWhyEppleyHannah2006} notices, this can only be achieved if the oscillators are at a temperature $T \lesssim \omega_0$; otherwise the increase would not be resolvable against thermal fluctuations.
On the other hand, due to the Hawking-Unruh effect\cite{hawkingParticleCreationBlack1975,unruhNotesBlackholeEvaporation1976} equipotential surfaces emit black body radiation\cite{wangSurfacesAwayHorizons2018} at a temperature proportional to the surface gravity. Hence, the oscillator cannot be at a temperature much lower than $T \sim (m + M_\text{tot})/R^2$ and we have
\begin{equation}
M_\text{tot}\gtrsim \frac{T R^2}{m^2} \sim \frac{1}{m} + \frac{M_\text{tot}}{m^2} \,.
\end{equation}
Considering the cases $M_\text{tot} > m$ and $M_\text{tot} < m$ separately, one finds that in both cases this implies $m \gtrsim 1$.
As a minimum requirement to violate the uncertainty relation with this type of experiment, even in principle, we need a particle mass of at least $m_P \approx \unit{2 \times 10^{-8}}{\kilogram}$. The uncertainty relation has not been confirmed experimentally for masses that large. There is also no theoretical reason to believe that the uncertainty relation must hold for all parameter regimes, especially if one attempts to fundamentally modify the principles of quantum mechanics as in most approaches for semiclassical gravity.

The deeper reason why many find in this a convincing argument against semiclassical gravity, or any classical-quantum coupling, is that it allows for a way to access quantum information, i.\,e.\ properties of a quantum state which are not measurable by any experiment in orthodox quantum mechanics. This tremendous deviation from established principles can indeed result in difficult to resolve paradoxes; retrieving quantum information, however, is not sufficient. As I will discuss in section \ref{sec:psi}, one needs to make use of nonlocal entanglement and attempt signalling with the retrievable quantum information in order to bring semiclassical gravity to bay.

Although Eppley and Hannah were the first to present a complete idea for a thought experiment, the argument is often attributed to the work of Bohr and Rosenfeld\cite{bohrZurFrageMessbarkeit1933}, allegedly demonstrating a consistency argument that would necessitate the quantization of the \emph{electromagnetic} field. The objection that their argument may not apply to gravity\cite{baymTwoslitDiffractionHighly2009} misses the point. As Rosenfeld himself points out\cite{rosenfeldQuantizationFields1963}, the argument does not even mandate quantization in the electromagnetic case. What Bohr and Rosenfeld actually show is that inconsistencies that arise in a rather naive---and already in its definition inconsistent---treatment, that mixes classical and quantum concepts, is resolved \emph{if} one treats the electromagnetic field as properly quantized. Notably, this is not an \emph{if and only if}.

\subsection{Causality violating acquisition of which-path information}\label{sec:subb}

Assume Alice wants to make use of the gravitational interaction to send a message to Bob. Alice has at her disposal a sufficiently large mass $M$ which she can move from an initial position $a_0$ into different positions, say $a_1$ and $a_2$. Bob is in possession of a test mass $m$. By monitoring the position of $m$ for some time $\tau$, Bob will be able to tell from the final position $b_1$ or $b_2$ the position of Alice's mass, opening a channel for communication.

Let $a_1 < a_2 < b_1 < b_2$ all be on the $x$-axis with $2 \Delta a = a_2 - a_1$ and $d \gg \Delta a$ the average distance between Alice and Bob. A multipole expansion of the Newtonian gravitational potential $\Phi(\vec r) = -\int \rmd^3 r' \rho(\vec r') \abs{\vec r - \vec r'}^{-1}$ at Bob's position around the average position of Alice's mass yields
\begin{equation}
\Phi_{1,2} = -\frac{M}{d} \pm \frac{D_a}{d^2} - \frac{Q_a}{d^3} \pm \frac{O_a}{d^4} + \dots  \,,
\end{equation}
where $D_a = M \Delta a$, $Q_a = M \Delta a^2$, $O_a = M \Delta a^3$ are the \emph{virtual} gravitational dipole, quadrupole, and octopole moments\footnote{Belenchia et al.\cite{belenchiaQuantumSuperpositionMassive2018} point out that, due to conservation of the center of mass, there is no dipole moment if one considers the entire system of Alice's mass and the surrounding lab. In the symmetric situation considered here, there is also no contribution to $\Delta b$ from the quadrupole\cite{rydvingGedankenExperimentsCompel2021}. For the further discussion, we assume that Alice and Bob are both located within a sufficiently large laboratory together, such that there is a dipole contribution. The generalization to situations where the leading order contributions stem from the quadrupole or octopole moments is straightforward.} associated with Alice's possible position choices (note that there is no real multipole in the classical case). Then we find $2 \Delta b = b_2 - b_1 \approx \tau^2 D_a / d^3$. If Bob can measure $\Delta b$ with a resolution $\delta$, then he can determine the state of Alice's mass in a time shorter than the travel time of a light signal from Alice to Bob, as long as $D_a > \delta d$.
Obviously, in classical physics there is no way to actually send a signal faster than light. In order to send a signal, Alice's state must \emph{change} and the consequences of this change will be transmitted to Bob in the form of gravitational waves which only travel at the speed of light. Note that Bob's role is entirely passive; he is the recipient of the signal and his actions have no influence on Alice's system in return.

In quantum mechanics, we are facing a slightly different situation\cite{mariExperimentsTestingMacroscopic2016}, as it is possible to find Alice's settings $a_1$ and $a_2$ in \emph{superposition}. Bob measuring $\Delta b$, on the other hand, is gaining information about the position of Alice's particle which, according to the complementarity principle in quantum mechanics, should decohere Alice's state. Contrary to the classical situation, Bob's action now does have a backwards influence on Alice's system.
This opens the possibility for Alice to determine whether or not Bob has performed his measurement and thereby allowing some message to be sent from Bob to Alice.
In order to detect if her state has decohered, Alice must perform some type of interference experiment which will take some time $T$. Faster-than-light signalling is possible as long as $T + \tau < d$.

Considering the finite speed at which changes in the gravitational potential propagate does not help the situation; Alice can have her state readily prepared long before Bob even starts thinking about performing his measurement. The signalling from Bob to Alice is due to the nonlocal entanglement of their respective quantum states, together with the quantum mechanical description of collapse as an instantaneous effect on the wave function everywhere. One may, of course, ask whether an instantaneous collapse is not immediately inconsistent with any relativistic theory, and in fact, a Lorentz invariant description in which the collapse happens (at least) instantaneously in every frame should have the rather odd property that the collapse propagates along the backwards light cone of a measurement event. Nonetheless, the (for all practical purposes) instantaneity of the collapse has been confirmed in experiments\cite{stefanovQuantumCorrelationsSpacelike2002}. For lack of a better alternative, we assume the usual quantum mechanical description to apply: that Bob's acquisition of which-path information decoheres Alice's state even for spacelike separations. Replacing the dynamical equations (i.\,e.\ the \schr\ equation) with corresponding relativistic dynamics (e.\,g.\ quantum field theory) does not avoid the possibility of faster-than-light signalling. 
Taking the dynamical character of the gravitational interaction into consideration, nevertheless, resolves the paradox. This has been detailed by Belenchia et al.\cite{belenchiaQuantumSuperpositionMassive2018} for the case of quantized gravity and I will reiterate their arguments in a more general fashion.

Let us first look at the constraints on Bob's resolution $\delta$. In order to observe the position of his test mass, Bob must in some way interact with it through the exchange of some particle with energy $E$ and wave length $\lambda$, for instance, a photon being scattered off the test mass and reaching Bob's eye. The resolution is limited by the wave length, $\delta > \lambda$. On the other hand, the resolution is limited by the scattering cross section which must be larger than the Schwarzschild radius $r_S = 2 E$ of the particle. We have $\delta \gtrsim 2 E \geq 2 p = 4 \pi / \lambda > 4 \pi / \delta$ which implies $\delta > 2 \sqrt{\pi} > 1$. The test mass position can only be determined up to Planck length precision, and with the considerations from above we find $D_a > d$ as the condition for faster-than-light signalling.

In order to perform her interference experiment, Alice must bring the two states in spatial superposition back to one location, eliminating her dipole moment in time $T$ by an acceleration $\ddot{D}_a \sim D_a/T^2$. With the Larmor formula for gravitoelectromagnetism\cite{clarkGaugeSymmetryGravitoelectromagnetism2000} one finds that this amounts to a total gravitational energy $E \sim \ddot{D}_a^2 T \sim D_a^2 / T^3$ being radiated away in form of gravitational waves. We can in principle gain which-path information from the emitted waves, resulting in a loss of coherence before Alice has the chance to finish her experiment. However, if some quantum system is used for the detection, this is only possible if the energy exceeds the threshold given by the time-energy uncertainty relation, $E T > 1$. Hence, Alice will be able to successfully perform her experiment as long as $D_a < T$. In order to send a signal faster than light, one then requires $D_a < T < d$, in contradiction to the requirement $D_a > d$ from the previous paragraph. We conclude that there is no possibility for faster-than-light signalling.

Note that we did not require explicitly that the gravitational field be quantized. We only need it to be capable of carrying which-path information. If, in a semiclassical theory, gravity carries no such information, Alice's state will not decohere, neither from the emission of radiation \emph{nor} from Bob's measurement. In this case, there is no problem of signalling in the first place.

Belenchia et al.\cite{belenchiaQuantumSuperpositionMassive2018} present the above argument only for the concrete example of pertubatively quantized gravity in analogy to quantum electrodynamics. The Planck length limit for Bob's resolution is then understood as a limit from vacuum fluctuations of the curvature tensor. The condition $D_a < T$, on the other hand, can be phrased as an emission of not even a single graviton of wave length $T$, which is not an essentially different criterion from the one given above based on the emission of classical gravitational waves. A different point of view is the requirement that interference fringes should be at least a Planck length apart in order to be detectable\cite{rydvingGedankenExperimentsCompel2021}, which results in similar restrictions.

The definiteness of the above analysis with regard to the impossibility of faster-than-light signalling can be doubted based on the assessment that many of the relations were only approximate. Hence, one may ask if in settings where they are close to being satisfied there could be possibilities to send a signal just above the speed of light, which would suffice for a claim of inconsistency. There are also ideas\cite{bekensteinTabletopSearchPlanck2012} which would allow to detect the mass center of a solid body with a precision $\delta < 1$, at least in principle, invalidating the arguments above which assume that Bob's test mass is a point like particle\footnote{Of course, one can simply postulate the Planck length as a fundamental limit for position measurements\cite{rydvingGedankenExperimentsCompel2021}.}. The essence of the argument, however, remains that it does not matter whether one quantizes the gravitational field or not; the limitations from vacuum fluctuations of curvature\cite{belenchiaQuantumSuperpositionMassive2018} are approximate and rely on the point particle property just as much as the classical arguments presented here. To the degree to which the former is evidence for consistency of quantized gravity, the latter should be regarded as evidence for the consistency of semiclassical gravity.

The thought experiment \emph{did} require both an instantaneous collapse of Alice's state upon Bob's measurement and the validity of the complementarity principle. Had we found some violation of causality, we could have attempted to amend it by allowing for deviations of either or both principles. The outcome of the analysis shows, however, that the thought experiment is perfectly causal without the need of any such fundamental changes.

\section{Causality paradox in psi-ontic semiclassical gravity}\label{sec:psi}

In the previous section, we learned that two of the commonly discussed paradoxes regarding semiclassical gravity are not paradoxical after all. First of all, they do not apply to the orthodox, $\psi$-ontic models, and second of all, even in the cases where they do apply, specifically for the hidden variable models, they are easily resolved.
We now address a more serious issue which arises (at least) for the $\psi$-ontic models.

Assume we have a pair of entangled spin-$\tfrac12$ particles, with Alice and Bob each in possession of one of these particles, hence having access to half of the entangled state
\begin{equation}
\ket{\Psi} = \alpha \ket{\uparrow}_a \otimes \ket{\downarrow}_b
+ \beta \ket{\downarrow}_a \otimes \ket{\uparrow}_b \,.
\end{equation}
If Alice performs a spin measurement, this state collapses with probabilities $\abs{\alpha}^2$ and $\abs{\beta}^2$, respectively, into one of the two summands, resulting in an ensemble with density matrix 
\begin{equation}\label{eqn:rhoc}
 \hat{\rho}_c = \abs{\alpha}^2 \ket{\uparrow \downarrow} \bra{\uparrow \downarrow\,}
 + \abs{\beta}^2 \ket{\downarrow \uparrow} \bra{\downarrow \uparrow\,} \,,
\end{equation}
where we write $\ket{\uparrow \downarrow} = \ket{\uparrow}_a \otimes \ket{\downarrow}_b$ etc.
If Alice does not perform the measurement, on the other hand, we find the density matrix of the pure state to be
\begin{equation}\label{eqn:rhop}
 \hat{\rho}_p = \hat{\rho}_c 
  + \alpha \beta^* \ket{\uparrow \downarrow} \bra{\downarrow \uparrow\,} 
 + \alpha^* \beta \ket{\downarrow \uparrow} \bra{\uparrow \downarrow\,} \,.
\end{equation}
Bob, only being able to perform a measurement on his part of the state, must trace out Alice's degrees of freedom. The interference terms in the pure density matrix \eqref{eqn:rhop} then vanish and one ends up with the same reduced density matrix
\begin{equation}
\hat{\rho}_b = \abs{\alpha}^2 \ket{\uparrow}_b \bra{\uparrow\,}_b + \abs{\beta}^2 \ket{\downarrow}_b \bra{\downarrow\,}_b 
\end{equation}
regardless of Alice's decision to (not) perform a measurement.

Introducing the basis labeling $\{\ket{i}\}_{i \in 1\dots 4} = \{\ket{\uparrow\uparrow}, \ket{\uparrow\downarrow}, \ket{\downarrow\uparrow}, \ket{\downarrow\downarrow}\}$, we can express any unitary time evolution by a unitary matrix such that $\ket{i} \to \sum_j U_{ij}(t) \ket{j}$, which induces an evolution law for the density matrices \eqref{eqn:rhoc} and \eqref{eqn:rhop}:
\begin{align}
\hat{\rho}_c &\to \sum_{i,j} \left(\abs{\alpha}^2 U_{2i} U_{2j}^* + \abs{\beta}^2 U_{3i} U_{3j}^* \right) \ket{i}\bra{j} \\
\hat{\rho}_p &\to \sum_{i,j} \left(\alpha U_{2i} + \beta U_{3i}\right)\left(\alpha^* U_{2j}^* + \beta^* U_{3j}^*\right) \ket{i}\bra{j} \,.
\end{align}
The difference between the reduced density matrices becomes
\begin{equation}
 \delta \hat{\rho}_b(t) = \Tr_a \left(\hat{\rho}_p - \hat{\rho}_c\right)
 = \alpha \beta^* \begin{pmatrix}
U_{21} U_{31}^* + U_{23} U_{33}^* & U_{21} U_{32}^* + U_{23} U_{34}^* \\
U_{22} U_{31}^* + U_{24} U_{33}^* & U_{22} U_{32}^* + U_{24} U_{34}^*
\end{pmatrix} + \text{adj.}
\end{equation}
This expression is generally nonzero. However, if the evolution affects only Bob's state and leaves Alice's unaltered, the matrix $U$ becomes block diagonal and we find $\delta \hat{\rho}_b(t) = 0$. Hence, Bob is unable to distinguish between the collapsed ensemble $\hat{\rho}_c$ and the pure state $\hat{\rho}_p$ by any local experiment.

Let us be more concrete and assume that Bob determines the spin by performing a Stern-Gerlach experiment, i.\,e.\ he subjects his particle to a magnetic field gradient for some short time $\tau_\text{acc}$ with a sign change after half that time, such that the spin state becomes entangled with the particles position, $\ket{x_1}$ for the spin state $\ket{\uparrow}_b$ and $\ket{x_2}$ for the spin state $\ket{\downarrow}_b$.
A local experiment at Bob's position then results in a time evolution
\begin{subequations}\begin{align}
\ket{\uparrow x_2} &\to \int \rmd x \, a_2(x) \ket{\uparrow x} \\
\ket{\downarrow x_1} &\to \int \rmd x \, a_1(x) \ket{\downarrow x} \\
\ket{\Psi} = \alpha \ket{\uparrow x_2} + \beta \ket{\downarrow x_1} &\to \int \rmd x \, \left(\alpha \widetilde{a}_2(x) \ket{\uparrow x} + \beta \widetilde{a}_1(x) \ket{\downarrow x} \right) \,.
\end{align}\end{subequations}
Bob's reduced density matrix in position space ends up to be
\begin{align}
\hat{\rho}_{c,b}(x,y) &= \abs{\alpha}^2 a_2(x) a_2^*(y) + \abs{\beta}^2 a_1(x) a_1^*(y)
\label{eqn:red-density-cb}\\
\hat{\rho}_{p,b}(x,y) &= \abs{\alpha}^2 \widetilde{a}_2(x) \widetilde{a}_2^*(y) + \abs{\beta}^2 \widetilde{a}_1(x) \widetilde{a}_1^*(y) \,,
\label{eqn:red-density-pb}
\end{align}
for the case of a collapsed wave function and the pure state $\ket{\Psi}$, respectively.
In standard quantum mechanics, the linearity of the time evolution law requires $\widetilde{a}_{1,2} = a_{1,2}$ and, hence, ensures that the density matrices \eqref{eqn:red-density-cb} and \eqref{eqn:red-density-pb} are identical.

In the orthodox semiclassical approach~\eqref{eqn:sce}, a generic state
\begin{equation}
 \ket{\chi} = \int \rmd x \, \left(\alpha \chi_\uparrow(x) \ket{\uparrow x} + \beta \chi_\downarrow(x) \ket{\downarrow x}\right)
\end{equation}
results in the mass density distribution
\begin{equation}
 \bra{\chi} \hat{m}(x) \ket{\chi} = m \braket{\chi}{x}\braket{x}{\chi}
 = m \abs{\alpha}^2 \abs{\chi_\uparrow(x)}^2 + m \abs{\beta}^2 \abs{\chi_\downarrow(x)}^2 \,.
\end{equation}
For the situation of interest, we find $\abs{\chi_\uparrow(x)}^2 \sim \delta(x-x_2)$ and $\abs{\chi_\downarrow(x)}^2 \sim \delta(x-x_1)$ and hence Newtonian potentials
\begin{subequations}\begin{align}
 V_2(x) &= -\frac{m^2}{\abs{x-x_2}} \\
 V_3(x) &= -\frac{m^2}{\abs{x-x_1}} \\
 V_\Psi(x) &= \abs{\alpha}^2 V_2(x) + \abs{\beta}^2 V_3(x)
\end{align}\end{subequations}
for the states $\ket{2} = \ket{\uparrow\downarrow} = \ket{\uparrow x_2}$, $\ket{3} = \ket{\downarrow\uparrow}= \ket{\downarrow x_1}$, and $\ket{\Psi} = \alpha \ket{2} + \beta \ket{3}$, respectively. Ignoring the free spreading of the wave function as well as the self-gravitational effects of these potentials on the wave function, the states $\ket{2}$ and $\ket{3}$ are unaffected, whereas the superposition state experiences a shift
\begin{equation}
 \ket{\Psi} \quad \to \quad \ket{\Psi}_t = \alpha \ket{\uparrow \widetilde{x}_2} + \beta \ket{\downarrow \widetilde{x}_1}
\end{equation}
with
\begin{equation}\label{eqn:deltax}
 \widetilde{x}_1 = x_1 - \abs{\alpha}^2 \delta x \,,\quad
\widetilde{x}_2 = x_2 + \abs{\beta}^2 \delta x \,,\quad
\delta x \approx \frac{m t^2}{2 \Delta x^2} \,,
\end{equation}
for $\Delta x = x_1 - x_2 > 0$, without loss of generality, and assuming $\delta x \ll \Delta x$. Hence, we have $a_i(x) \approx \delta(x - x_i) \neq \widetilde{a}_i(x) \approx \delta(x - \widetilde{x}_i)$ and the density matrices \eqref{eqn:red-density-cb} and \eqref{eqn:red-density-pb} become distinguishable. They predict different outcomes for position measurements at Bob's particle: $x_1$ or $x_2$ versus $\widetilde{x}_1$ or $\widetilde{x}_2$ (in both cases with probabilities $\abs{\alpha}^2$ and $\abs{\beta}^2$). Although we focus on the result of the semiclassical Einstein equations here, other $\psi$-ontic models face the same issue. The gravitational potential is determined by the wave function, rendering the Schrödinger evolution nonlinear in the wave function.

It has been argued\cite{gisinStochasticQuantumDynamics1989,bahramiSchrodingerNewtonEquationIts2014} that this distinguishability can be exploited to violate causality, because the reduction from \eqref{eqn:red-density-cb} to \eqref{eqn:red-density-pb} happens instantaneously upon measurement in standard quantum mechanics. In fact, this argument that causality requires a linear evolution of the density matrix, is the very basis upon which the theoretical formalism of collapse models is founded. There are good reasons to question the conclusiveness of this claim. For instance, Kent\cite{kentNonlinearitySuperluminality2005,kentTestingQuantumGravity2021} has shown that a description of measurement based on the ``local state'' of a particle, i.\,e.\ its reduced density matrix conditioned on all the measurement outcomes in the past light cone, allows for nonlinear evolution without the possibility to signal faster than light.

On the other hand, even if one \emph{does} believe that distinguishability between \eqref{eqn:red-density-cb} and \eqref{eqn:red-density-pb} poses a problem one may ask if the difference can ever be observed, even in principle, in any sort of experiment that would allow for faster-than-light signalling. Note that the possibility to signal is not (at least not entirely) due to the nonlinear gravitational interaction, it is due to the projective spin measurement and the induced instantaneous, nonlocal collapse of the wave function. Hence, the details of the collapse mechanism are likely to matter.

In order to actually resolve the position shift $\delta x$, it must be larger than the free spreading of the wave function due to position-momentum uncertainty:
\begin{equation}\label{eqn:inequal-pos-mom}
 \delta x^4 \gtrsim \frac{t^2}{m^2} 
 \sim \frac{\delta x \, \Delta x^2}{m^3}
 \quad \Rightarrow \quad
 m^3 \gtrsim \frac{\Delta x^2}{\delta x^3}
 \gg \frac{1}{\delta x} \,.
\end{equation}
This criterion can of course be satisfied simply by choosing a sufficiently large mass.
However, if we account for a dynamical collapse of the wave function, a larger mass usually implies a faster collapse and we must take care that the superposition is maintained throughout the entire time $t$ of the experiment.

Instead of a precise dynamical law, we consider some prototype of a collapse dynamics which resembles the ideas of Di\'osi\cite{diosiModelsUniversalReduction1989} and Penrose\cite{penroseGravityRoleQuantum1996}: whenever the wave function becomes wider than some collapse radius $r_c$, it collapses towards a position eigenstate at a rate determined by the time-energy uncertainty, $\tau E \sim 1$, where $E \sim m^2/r_c$ is the gravitational self-energy of the superposition state of size $r_c$. The radius $r_c$ is to be considered a free parameter---in Di\'osi's model it is a cut-off required to avoid divergences from localized mass densities, although we can also simply take it as a proportionality constant between the collapse rate and the squared mass, or even a function of mass itself. The collapse time $\tau$, on the other hand, follows from Penrose's argument that the uncertainty for the generators of time translation between the two spacetimes belonging to two classical states in superposition can be associated with the gravitational self-energy in precisely this way.
If, then, we require that the superposition must be maintained throughout the experiment, i.\,e.\ $t < \tau$, we have
\begin{equation}\label{eqn:rc-limit}
r_c \sim m^2 \tau \gtrsim m^2 t \sim \sqrt{m^3 \delta x \, \Delta x^2}
\gtrsim \frac{\Delta x^2}{\delta x} \gg \delta x \,,
\end{equation}
where we used the inequality \eqref{eqn:inequal-pos-mom} in the second to last and $\Delta x \gg \delta x$ in the last step. In conclusion, we can only observe a shift $\delta x$ that is \emph{below} the collapse radius $r_c$.
How small can $r_c$ be? The strongest experimental constraints stem from levitated nanoparticles\cite{delicCoolingLevitatedNanoparticle2020} which require $r_c \gtrsim \unit{10^{-15}}{\meter}$ for masses of $m \sim \unit{10^{-17}}{\kilogram}$.
If $r_c$ was in fact of this order of magnitude, according to the position-momentum uncertainty relation~\eqref{eqn:inequal-pos-mom} we would require a mass of at least $\unit{10^{-15}}{\kilogram}$ whose center-of-mass position we would need to resolve with femtometer precision.

\begin{figure}
 \centering
 \includegraphics[scale=0.6]{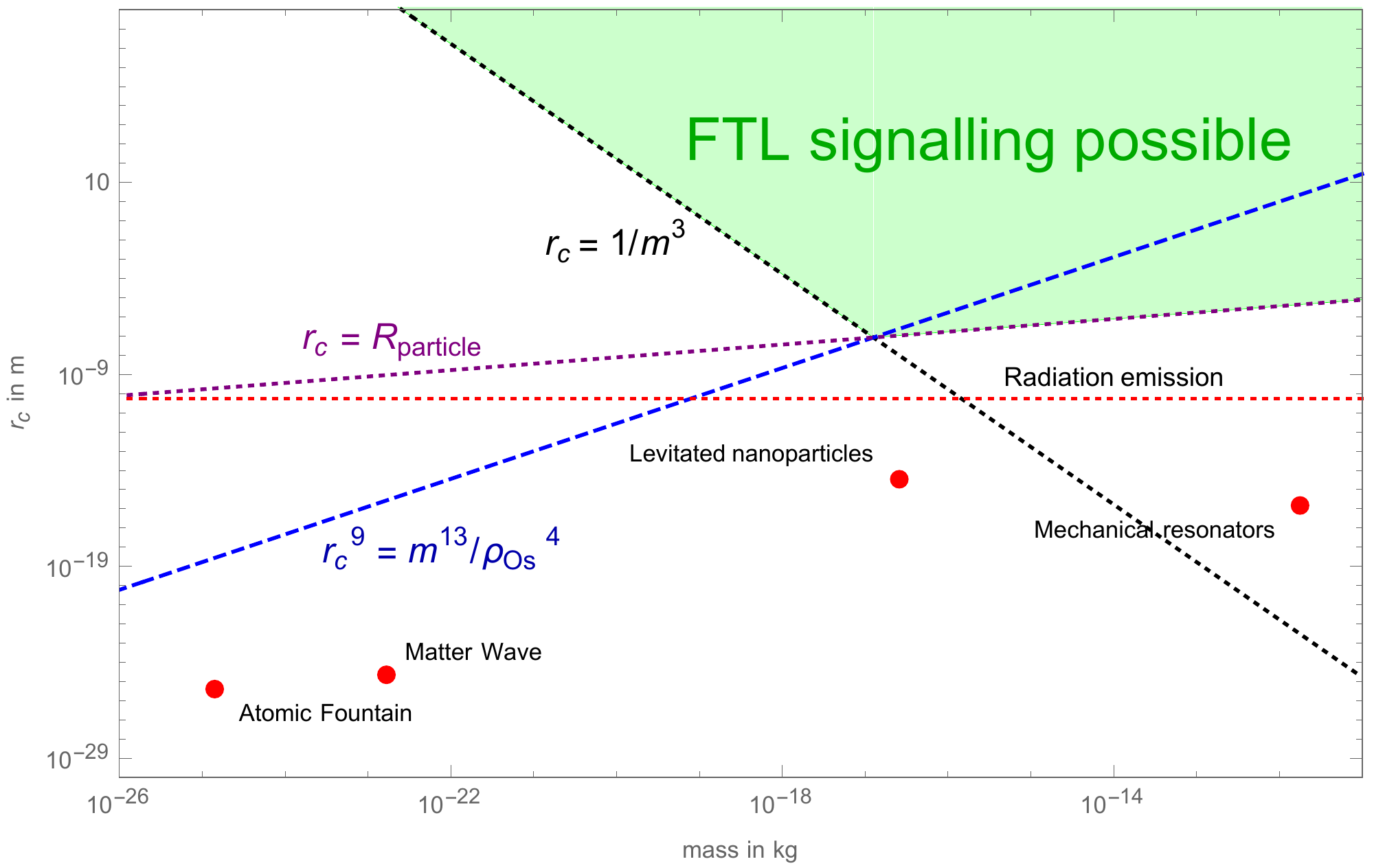}
 \caption{Exclusion plot for possible values of $r_c$. The dotted black line shows the curve $r_c = m^{-3}$, below which no detection of $\delta x$ is possible. The dashed blue line shows the curve $r_c^9 = m^{13}/\rho^4$ for the densest known element, osmium, below which no detection is possible for superpositions smaller than the particle radius, i.\,e.\ below the purple line. Any---possibly mass dependent---value of $r_c$ below the green shaded area would not allow detection before collapse and prevent faster-than-light signalling. The red dots show lower limits on $r_c$ from atomic fountain\cite{sugarbakerEnhancedAtomInterferometer2013}, matter wave\cite{eibenbergerMatterWaveInterference2013}, nanoparticle\cite{delicCoolingLevitatedNanoparticle2020}, and mechanical resonator\cite{oconnellQuantumGroundState2010} experiments, respectively. Neutron interferometry experiments\cite{colellaObservationGravitationallyInduced1975} would be in the far bottom left corner, excluded for better legibility. The red dotted line shows the recent limit on $r_c$ in the Di\'osi-Penrose collapse model\cite{donadiUndergroundTestGravityrelated2021}.}
 \label{fig:plot}
\end{figure}

Although the experiment is obviously difficult, we are not interested in a concrete realization. What matters is whether it is possible \emph{in principle}. Comparing equations \eqref{eqn:inequal-pos-mom} and \eqref{eqn:rc-limit}, we find that $r_c \gg 1/m^3$ must hold, in order to be able to acquire the necessary position shift due to semiclassical gravity before the wave function collapses. This implies that one needs a large mass for faster-than-light signalling, yet, quantum superpositions of large masses have been demonstrated and, in fact, mechanical resonators\cite{oconnellQuantumGroundState2010} achieve values which would lie above the $r_c \sim 1/m^3$ threshold. However, for massive particles the superposition size will generally be within the particle radius, $\Delta x < R$, and we must modify equation \eqref{eqn:deltax} to reflect the gravitational force between two overlapping spheres: $\delta x \sim \rho \,\Delta x \,t^2$ for mass density $\rho$. Equations \eqref{eqn:inequal-pos-mom} and \eqref{eqn:rc-limit} then read
\begin{align}
\delta x^9 &\gtrsim \frac{t^6}{m^6 \, \delta x^3} 
\sim \frac{1}{m^6 \, \rho^3 \, \Delta x^3} 
> \frac{1}{m^6 \, \rho^3 \, R^3} \sim \frac{1}{m^7 \rho^2}\\
r_c^9 &\sim m^{18} \tau^9 \gtrsim \sqrt{\frac{m^{36} \delta x^9}{\rho^9 \Delta x^9}} 
> \sqrt{\frac{m^{33} \delta x^9}{\rho^6}} 
> \frac{m^{13}}{\rho^4} \,.\label{eqn:rc-massive}
\end{align}
The two conditions \eqref{eqn:rc-limit} and \eqref{eqn:rc-massive} intersect at the particle radius $r_c = R$.

The picture we are left with is illustrated in figure~\ref{fig:plot} as an exclusion plot for $r_c$ in terms of the mass. The dotted purple line shows the values for which $r_c$ equals the particle radius for an osmium\footnote{We choose osmium because it is the densest known element. Note that the effect of the density on the result is marginal, as long as one does not consider densities many orders of magnitude above those naturally occurring in solid bodies.} sphere of a given mass. For values above that line, the detection of $\delta x$ is limited by the criterion \eqref{eqn:rc-limit}, corresponding to the dotted black line; for values below, it is limited by \eqref{eqn:rc-massive}, the dashed blue line. Since the condition \eqref{eqn:rc-massive} poses a stricter criterion than $r_c = R$, the actual limitation on detection of $\delta x$ is set by the latter. Detecting the shift $\delta x$ and, therefore, being able to send a faster-than-light signal is possible only for values of $r_c$ in the green shaded area.\footnote{Note that, although the criteria \eqref{eqn:rc-limit} and \eqref{eqn:rc-massive} are satisfied in the area between the dashed blue and dotted purple lines for small masses below the dotted black line, in this area one would require a contradictory spatial resolution $\delta x > \Delta x$.} The plot also shows the lower limits put on $r_c$ by certain experiments. Any value $r_c$ or function $r_c(m)$ between the red dots and the green shaded area induces a collapse that is fast enough to prevent faster-than-light signalling and is compatible with observation. Levitated nanoparticles are possibly the preferable choice of experiment to exclude $r_c$ values that can avoid faster-than-light signalling. The recent limit\cite{donadiUndergroundTestGravityrelated2021} on the free parameter $r_c \gtrsim 0.54 \times 10^{-10}~$m of the Di\'osi-Penrose model\cite{diosiModelsUniversalReduction1989} from underground tests of radiation emission is plotted as a dotted red line. As a limit on the here proposed type of collapse it must be taken with caution, because the experiment did not involve actual spatial superposition states of that size; and even this result cannot exclude non-signalling semiclassical gravity.

The arguments presented here were specifically tailored to the gravitational coupling via the semiclassical Einstein equations~\eqref{eqn:sce}; one would need to repeat a similar analysis for other models to assure their consistency. It may also be possible to construct experiments other than the one described here in order to exploit the nonlinear evolution for signalling. Be that as it may, based on the current state of observation, there is no reason to believe that semiclassical gravity \emph{necessarily} violates causality as long as it is accompanied by some collapse mechanism---as mandated by the Page and Geilker experiment.

\section{Summary and discussion}\label{sec:discussion}

I have reviewed the consistency arguments that are most commonly raised against semiclassical gravity and have shown how all of them can be avoided if one accepts that a semiclassical theory of gravitation does not only require a coupling mechanism for quantum fields to spacetime curvature, but it must also provide a dynamical description of wave function collapse.
The arguments for consistency presented here were not based on any elaborate model for wave function collapse but rather on ad-hoc expectations on some basic features of such a model. Whether a consistent, fully relativistic model compatible with general relativity and the semiclassical coupling exists, remains an open problem. Nonetheless, the discussion shows that there is nothing preventing a fully consistent theory in principle.

With the necessity of collapse in mind, there are three approaches one may take in order to define a quantum matter-gravity coupling: the $\psi$-ontic one, taking the wave function as an element of physical reality responsible of sourcing spacetime curvature, the hidden variable approach, postulating some novel, non-quantum degree of freedom, or stochastic models aiming at some ``minimally invasive'' way of modifying quantum mechanics.

In that last category of stochastic semiclassical gravity, the model by Tilloy and Di\'osi\cite{tilloySourcingSemiclassicalGravity2016} represents a natural way to source gravity \emph{if} one believes in the presence of a stochastic collapse with linear master equation as in collapse models\cite{bassiModelsWavefunctionCollapse2013}. Besides the somewhat uncalled for occurrence of said collapse, which assumes the introduction of some non-quantum stochastic field, the main challenge is the generalization to a relativistic model.

Oppenheim's suggestion\cite{oppenheimPostquantumTheoryClassical2021} for a ``post-quantum'' semiclassical theory is formulated in a fully general relativistic fashion. However, as a stochastic model that describes ensembles of 3-manifolds rather than a single classical spacetime it raises the question if there is an underlying microscopic theory---as in classical statistical physics. It is also not entirely clear whether it can avoid the paradoxes of semiclassical gravity, as the full equations of motion are nonlinear and only become linear for matter after tracing out gravitational degrees of freedom, or the difficulties with self-energy and renormalization at high energies which render perturbative quantum gravity inconsistent.

In the light of these arguments, the $\psi$-ontic models---first and foremost the M\o{}ller-Rosenfeld model---remain an interesting possibility despite breaking with many established concepts of quantum theory. One question to devote oneself to is the cause of the necessary dynamical wave function collapse. Although the collapse could simply be caused by an additional, external mechanism, more convincing would be an explanation within semiclassical gravity itself.
The ingredients needed for a collapse are nonlinearity and stochasticity, of which the former is readily included in semiclassical gravity. The equations of motion of a semiclassical theory, on the other hand, are by default deterministic. The crucial question, therefore, is whether internal sources of randomness---for instance a random distribution of dark matter or a stochastic gravitational wave background---can provide boundary conditions that would result in stochastic behavior compatible with Born rule probabilities.

\begin{acknowledgments}
I gratefully acknowledge funding by the Volkswagen Foundation.
\end{acknowledgments}

\section*{Data Availability}
The data that support the findings of this study are available from the corresponding author upon reasonable request.

\section*{Author Declarations}
The author has no conflicts to disclose.

%

\end{document}